\documentstyle[pra,aps,preprint]{revtex}
\begin{document}

\title{ON THE FOUNDATIONS OF\\ THE TWO MEASURES FIELD THEORY}

\author
{E. I. Guendelman \thanks{guendel@bgu.ac.il} and A.  B. Kaganovich
\thanks{alexk@bgu.ac.il}}
\address{Physics Department, Ben Gurion University of the Negev, Beer
Sheva 84105, Israel}

             \maketitle

\begin{abstract}
Two Measures Field Theory (TMT) uses both the Riemannian volume
element $\sqrt{-g}d^4x$ and a new one $\Phi d^4x$ where the new
measure of integration $\Phi$ can be  build of four scalar fields.
Arguments in favor of TMT, both from the point of view of first
principles and from the TMT results are summarized. Possible
origin of the TMT and symmetries that protect the structure of TMT
are reviewed. It appears that four measure scalar fields treated
as "physical coordinates" allow to define local observables in
quantum gravity. The resolution of the old cosmological constant
problem as a possible direct consequence of the TMT structure is
discussed. Other applications of TMT to cosmology and particle
physics are also mentioned.
\end{abstract}

\section{Main ideas of the Two Measures Field Theory}

TMT is a generally coordinate invariant theory where {\it all the
difference from the standard field theory in curved space-time
consists only of the following three additional assumptions}:

1. The first assumption is the hypothesis that the effective
action at the energies below the Planck scale has to be of the
form\cite{GK1}-\cite{HIS}
\begin{equation}
    S = \int L_{1}\Phi d^{4}x +\int L_{2}\sqrt{-g}d^{4}x
\label{S}
\end{equation}
 including two Lagrangians $ L_{1}$ and $L_{2}$ and two
measures of integration $\sqrt{-g}$ and $\Phi$ or, equivalently,
two volume elements $\Phi d^{4}x$ and $\sqrt{-g}d^{4}x$
respectively. One is the usual measure of integration $\sqrt{-g}$
in the 4-dimensional space-time manifold equipped with the metric
 $g_{\mu\nu}$.
Another  is the new measure of integration $\Phi$ in the same
4-dimensional space-time manifold. The measure  $\Phi$ being  a
scalar density and a total derivative may be defined\footnote{For
applications of the measure $\Phi$ in string and brane theories
see Ref.\cite{Mstring}.}
\begin{itemize}

\item
either by means of  four scalar fields $\varphi_{a}$ ($a=1,2,3,4$)
\begin{equation}
\Phi
=\varepsilon^{\mu\nu\alpha\beta}\varepsilon_{abcd}\partial_{\mu}\varphi_{a}
\partial_{\nu}\varphi_{b}\partial_{\alpha}\varphi_{c}
\partial_{\beta}\varphi_{d}.
\label{Phi}
\end{equation}

\item
or by means of a totally antisymmetric three index field
$A_{\alpha\beta\gamma}$
\begin{equation}
\Phi
=\varepsilon^{\mu\nu\alpha\beta}\partial_{\mu}A_{\nu\alpha\beta}.
\label{Aabg}
\end{equation}
\end{itemize}

To provide parity conservation in the case given by Eq.(\ref{Phi})
one can choose for example one of $\varphi_{a}$'s to be a
pseudoscalar; in the case given by Eq.(\ref{Aabg}) we must choose
$A_{\alpha\beta\gamma}$ to have negative parity.  A special case
of the structure (\ref{S}) with definition (\ref{Aabg}) has been
recently discussed in Ref.\cite{hodge} in applications to
supergravity and the cosmological constant problem.

2. It is assumed that the Lagrangians $ L_{1}$ and $L_{2}$ are
functions of all matter fields, the metric, the connection  (or
spin-connection )
 but not of the
"measure fields" ($\varphi_{a}$ or $A_{\alpha\beta\gamma}$).

3. Important feature of TMT that is responsible for many
interesting and desirable results of the field theory models
studied so far\cite{GK1}-\cite{GK6}
 consists of the assumption that all fields, including
also metric, connection (or vierbein and spin-connection) and the
{\it measure fields} ($\varphi_{a}$ or $A_{\alpha\beta\gamma}$)
are independent dynamical variables. All the relations between
them are results of equations of motion.  In particular, the
independence of the metric and the connection means that we
proceed in the first order formalism and the relation between
connection and metric is not necessarily according to Riemannian
geometry.

We want to stress again that except for the listed three
assumptions in our TMT models we do not make any changes as
compared with principles of the standard field theory in curved
space-time. In other words, all the freedom in constructing
different models in the framework of TMT consists of the choice of
the concrete matter content and the Lagrangians $ L_{1}$ and
$L_{2}$ that is quite similar to the standard field theory.

Since $\Phi$ is a total derivative, a shift of $L_{1}$ by a
constant, $L_{1}\rightarrow L_{1}+const$, has no effect on the
equations of motion. Similar shift of $L_{2}$ would lead to the
change of the constant part of the Lagrangian coupled to the
volume element $\sqrt{-g}d^{4}x $. In the standard GR, this
constant term is the cosmological constant. However in TMT the
relation between the constant
 term of $L_{2}$ and the physical cosmological constant is very non
trivial (see \cite{GK3}-\cite{K}).

In the case of the definition of $\Phi$ by means of
Eq.(\ref{Phi}), varying the measure fields $\varphi_{a}$, we get
\begin{equation}
B^{\mu}_{a}\partial_{\mu}L_{1}=0  \quad where \quad
B^{\mu}_{a}=\varepsilon^{\mu\nu\alpha\beta}\varepsilon_{abcd}
\partial_{\nu}\varphi_{b}\partial_{\alpha}\varphi_{c}
\partial_{\beta}\varphi_{d}.\label{varphiB}
\end{equation}
Since $Det (B^{\mu}_{a}) = \frac{4^{-4}}{4!}\Phi^{3}$ it follows
that if $\Phi\neq 0$,
\begin{equation}
 L_{1}=sM^{4} =const
\label{varphi}
\end{equation}
where $s=\pm 1$ and $M$ is a constant of integration with the
dimension of mass.

In the case of the definition (\ref{Aabg}), variation of
$A_{\alpha\beta\gamma}$ yields
\begin{equation}
\varepsilon^{\mu\nu\alpha\beta}\partial_{\mu}L_{1}=0, \label{AdL1}
\end{equation}
that implies Eq.(\ref{varphi}) without the  condition $\Phi\neq 0$
needed in the model with four scalar fields $\varphi_{a}$.

 One should notice
 {\it the very important differences of
TMT from scalar-tensor theories with nonminimal coupling}: \\
 a) In general, the Lagrangian density $L_{1}$ (coupled to the measure
$\Phi$) may contain not only the scalar curvature term (or more
general gravity term) but also all possible matter fields terms.
This means that TMT modifies in general both the gravitational
sector  and the matter sector; b) If the field $\Phi$ were the
fundamental (non composite) one then instead of (\ref{varphi}),
the variation of $\Phi$ would result in the equation $L_{1}=0$ and
therefore the dimensionfull integration constant $M^4$ would not
appear in the theory.

Applying the Palatini formalism in TMT one can show (see for
example \cite{GK3},\cite{HIS})  that the resulting relation
between metric and  connection includes also the gradient of the
ratio of the two measures
\begin{equation}
\zeta \equiv\frac{\Phi}{\sqrt{-g}} \label{zeta}
\end{equation}
which is a scalar field. The gravity and matter field equations
obtained by means of the first order formalism contain both
$\zeta$ and its gradient. It turns out that at least at the
classical level, the measure fields affect the theory only through
the scalar field $\zeta$.

The consistency condition of equations of motion has the form of a
constraint which determines $\zeta (x)$ as a function of matter
fields. The surprising feature of the theory is
 that neither Newton constant nor curvature appear in this constraint
which means that the {\it geometrical scalar field} $\zeta (x)$
{\it is determined by the matter fields configuration}  locally
and straightforward (that is without gravitational interaction).

By an appropriate change of the dynamical variables which includes
a conformal transformation of the metric, one can formulate the
theory in a Riemannian (or Riemann-Cartan) space-time. The
corresponding conformal frame we call "the Einstein frame". The
big advantage of TMT is that in the very wide class of models,
{\it the gravity and all matter fields equations of motion take
canonical GR form in the Einstein frame}.
 All the novelty of TMT in the Einstein frame as compared
with the standard GR is revealed only
 in an unusual structure of the scalar fields
effective potential (produced in the Einstein frame), masses of
fermions  and their interactions with scalar fields as well as in
the unusual structure of fermion contributions to the
energy-momentum tensor: all these quantities appear to be $\zeta$
dependent. This is why the scalar field $\zeta (x)$ determined by
the constraint as a function of matter fields, has a key role in
dynamics of TMT models.

\section{Possible origin of TMT from low energy limit
of the brane-world scenario}

As for the possible origin of the modified measure one can think
of the following arguments. Let us start by noticing that the
modified measure part of the action (\ref{S}), $\int L_{1}\Phi
d^{4}x$, in the case $ L_{1}=const$ becomes a topological
contribution to the action, since $\Phi$ is a total derivative. It
is very interesting that this structure can be obtained by
considering the topological theory which results from studying
"space-time filling branes", Refs.\cite{Zai}-\cite{Aka}.

Following for example Ref.\cite{Flor}, the Nambu-Goto action of a
3-brane embedded in a $4$-dimensional space-time\footnote{It
should be pointed out that string and brane theories can be
formulated with the use of a modified measure\cite{Mstring} giving
some new interesting results. In the spirit of the interpretation
developed in this section, such strings or branes could be
regarded as extended objects moving in an embedding extended
object of the same dimensionality.}
\begin{equation}
S_{NG}=T\int d^{4}x\sqrt{|\det (\partial_{\mu}\varphi^{a}
\partial_{\nu}\varphi^{b}G_{ab}(\varphi) |}
\label{NG}
\end{equation}
(where $G_{ab}(\varphi)$ is the metric of the embedding space), is
reduced to
\begin{equation}
\frac{T}{4!}\int d^{4}x
\varepsilon^{\mu\nu\alpha\beta}\varepsilon_{abcd}\partial_{\mu}\varphi_{a}
\partial_{\nu}\varphi_{b}\partial_{\alpha}\varphi_{c}
\partial_{\beta}\varphi_{d}
\sqrt{|\det G_{ab}(\varphi^{a})|}= \frac{T}{4!}\int \Phi
\sqrt{|\det G_{ab}(\varphi^{a})|}d^{4}x. \label{Sreduced}
\end{equation}
 where $\Phi$ is determined by Eq.(\ref{Phi}).
The Euler-Lagrange equations for $\varphi^{a}$ are identities
which is the result of the fact\cite{Flor} that the action
(\ref{Sreduced}) is topological.

This is the action of a "pure" space-filling brane governed just
by the identically constant brane tension. Let us now assume that
the 3-brane is equipped with its own metric $g_{\mu\nu}$ and
connection. If we want to describe a
 space-filling brane where gravity and matter are included
by means of the Lagrangian $ L_{1}=
L_{1}(g_{\mu\nu},connection,matter fields)$, we are led to the
following form of the brane contribution to the action
\begin{equation}
S_{1}= \int L_{1}\Phi \sqrt{|\det G_{ab}(\varphi^{a})|}d^{4}x
\label{S-red-NG-L1}
\end{equation}
Recall that $\sqrt{|\det G_{ab}(\varphi^{a})|}$ is a brane scalar.

If $ L_{1}$ is not identically a constant (we have seen above that
it may become a constant on the "mass shell", i.e. when equations
of motion are satisfied), then we are not talking any more of a
topological contribution to the action. Assuming again that $
L_{1}$ is $\varphi^{a}$ independent, it is easy to check that in
spite of emergence of an additional factor $\sqrt{|\det
G_{ab}(\varphi^{a})|}$ in Eq.(\ref{S-red-NG-L1}) as compared with
the first term in Eq.(\ref{S}), variation of $\varphi^{a}$ yields
exactly the same equation (\ref{varphi}) if $\Phi\neq 0$.
Moreover, all other equations of motion of the two measures theory
remains unchanged. The only effect of this additional factor
consists in the redefinition of the scalar field $\zeta$,
Eq.(\ref{zeta}), where $\sqrt{|\det G_{ab}(\varphi^{a})|}$ emerges
as an additional factor. Notice that the same results are obtained
if instead of $\sqrt{|\det G_{ab}(\varphi^{a})|}$ in
Eq.(\ref{S-red-NG-L1}) there will be arbitrary function of
$\varphi^{a}$.

 Continuing  discussion of the general
structure of the action in TMT one of course may ask: why 3-brane
moving only in $3+1$ dimensional embedding space-time? This can be
obtained also starting from higher dimensional "brane-world
scenarios".  Indeed, let us consider for example
 a 3-brane evolving in an embedding 5-dimensional space-time with
\begin{equation}
ds^{2}=G_{AB}dx^{A}dx^{B}=f(y)\hat{g}_{\mu\nu}(x^{\alpha})dx^{\mu}dx^{\nu}+
\gamma^{2}(y)dy^{2}, \quad  A,B=0,1,...4; \quad \mu,\nu =0,1,2,3.
\label{interval}
\end{equation}
Assuming that it is possible to ignore the motion of the brane in
the extra dimension, i.e. studying the brane with a fixed position
in extra dimension, $y=const$, one can repeat the above arguments
(starting with the Nambu-Goto action)
 where one needs to use only
four functions $\varphi^{a}(x^{\mu})$ which together with the
fifth ($x^{\mu}$ independent) component along the axis $y$
constitute the
 5-vector describing the embedding of our brane;
$L_{1}$ is again the $\varphi^{a}$ independent Lagrangian of the
gravity\footnote{Notice that connection coefficients of our
3-brane are those where indexes run only from zero to three and
these components do not suffer from discontinuities across the
brane\cite{Israel}} and matter on the brane.
 As a result we obtain
exactly the same effective action as in Eq.(\ref{S-red-NG-L1})
which describes a brane moving in the hypersurface $y=const$ of
the 5-dimensional space-time.

The brane theory action contains also a piece coming from the bulk
dynamics. We will assume that gravity and matter exist also in the
bulk where their action can be written in the form
\begin{equation}
S_{bulk}=\int\sqrt{-\hat{g}}f^{2}(y)\gamma(y) L_{bulk}d^{4}xdy
\label{s-bulk}
\end{equation}
where $\hat{g}\equiv det(\hat{g}_{\mu\nu})$ and $L_{bulk}$ is the
Lagrangian of the gravity and matter in the bulk. We are not
interested in the dynamics in the hole bulk but rather in the
effect of the bulk on the 4-dimensional dynamics. For this purpose
one can integrate out the perpendicular coordinate $y$ in the
action (\ref{s-bulk}). One can think of this integration
 in a spirit of  the procedure known as
averaging (for the case of compact extra dimensions see for
example Ref.\cite{S.Weinberg}).  We do not perform this
integration explicitly\footnote{The correspondent calculations
must take into account the discontinuity constraints.} here but we
expect that the resulting averaged contribution of the bulk
dynamics to the 4-dimensional action one can write down in the
form
\begin{equation}
S_{2}=\int\sqrt{-\hat{g}} L_{2}(\hat{g}_{\mu\nu}, connection,
matter fields)d^{4}x \label{s-brane-from bulk}
\end{equation}
that we would like to use as the
 second term of the postulated in Eq.(\ref{S})
general form of the TMT action in four dimensions.

Notice however that the geometrical objects in these two actions
may not be identical. The simplest assumption would be of course
to take $\hat{g}_{\mu\nu}\equiv g_{\mu\nu}$ (and also coinciding
connections). One may allow for the case $\hat{g}_{\mu\nu}\neq
g_{\mu\nu}$ nevertheless. In fact, brane theory allows naturally
bimetric theories\cite{Damour-Kogan} even if one starts with a
single bulk metric. In our case this can be due to the fact that
the metric at the brane and its average value may be different.
The bimetric theories\cite{Salam} give in any case only one
massless linear combination of the two metrics which one can
identify as long distance gravity. Connections  at the brane and
its average value may be different as well. But we expect this
difference to be small due to the continuity of the relevant
connection coefficients (see footnote 3). Therefore performing the
integration over extra dimension $y$ we are left with just one
independent connection which is important for TMT where the first
order formalism is supposed to be one of the basic principles.

\section{SYMMETRIES}

\subsection{Volume preserving diffeomorphissms}
The volume element $\Phi d^4x$ is invariant under the area
preserving diffeomorphisms\cite{GK1}, i. e. internal
transformations in the $\varphi_{a}$ space
\begin{equation}
\varphi_{a}\rightarrow\varphi_{a}^{\prime}=\varphi_{a}^{\prime}
(\varphi_{b}), \quad a,b=1,2,3,4
        \label{GNP4}
\end{equation}
which satisfy the area preserving condition
\begin{equation}
\epsilon_{abcd}
\frac{\partial\varphi_{a}^{\prime}}{\partial\varphi_{k}}
             \frac{\partial\varphi_{b}^{\prime}}{\partial\varphi_{l}}
             \frac{\partial\varphi_{c}^{\prime}}{\partial\varphi_{m}}
             \frac{\partial \varphi_{d}^{\prime}}{\partial\varphi_{n}}
 =\epsilon_{klmn}.
\label{vol-pres}
\end{equation}

\subsection{Symmetries related to the general structure of TMT}
In such a case, i.e. when the measure fields  enter in the theory
only via the measure $\Phi$,
  the action (\ref{S}) respects
 (up to an integral of a total
divergence) the infinite dimensional group of shifts\cite{GK3} of
the measure fields $\varphi_{a}$. In the case given by
Eq.(\ref{Phi}) these symmetry transformations have the form
\begin{equation}
\varphi_{a}\rightarrow\varphi_{a}+f_{a}(L_{1}),
\label{inf-dim-symm}
\end{equation}
 where
$f_{a}(L_{1})$ are arbitrary functions of  $L_{1}$ (see details in
Ref.\cite{GK3}); in the case given by Eq.(\ref{Aabg}) they read
$A_{\alpha\beta\gamma}\rightarrow
A_{\alpha\beta\gamma}+\varepsilon_{\mu\alpha\beta\gamma}
f^{\mu}(L_{1})$ where $f^{\mu}(L_{1})$ are four arbitrary
functions of $L_{1}$ and $\varepsilon_{\mu\alpha\beta\gamma}$ is
numerically the same as $\varepsilon^{\mu\alpha\beta\gamma}$. One
can hope that this symmetry should prevent emergence of a measure
fields dependence in $ L_{1}$ and $L_{2}$ after quantum effects
are taken into account.

\section{Spontaneously broken global scale invariance without
Goldstone boson}

Let us consider a model where the action (\ref{S}) is invariant
under symmetry transformations in such a way that the measure
fields $\varphi_a$ (or $A_{\alpha\beta\gamma}$) participate in the
transformations as well. The matter content of our model includes
the scalar field $\phi$, two  fermion fields ( neutrino $N$ and
 electron $E$) and electromagnetic field $A_{\mu}$. We allow
in both $L_{1}$ and $L_{2}$ all the usual terms considered in
standard field theory models in curved space-time. To give an
example, we present here a model\cite{GK6} where keeping the
general structure of Eq.(\ref{S}) it is convenient to represent
the action in the following form:
\begin{eqnarray}
&S&=\int d^{4}x e^{\alpha\phi /M_{p}}(\Phi +b\sqrt{-g})
\left[-\frac{1}{\kappa}R(\omega ,e) +
\frac{1}{2}g^{\mu\nu}\phi_{,\mu}\phi_{,\nu}\right] -\int d^{4}x
e^{2\alpha\phi /M_{p}}[\Phi V_{1} +\sqrt{-g}V_{2}]
\nonumber\\
&&  -\int
d^{4}x\sqrt{-g}\frac{1}{4}g^{\alpha\beta}g^{\mu\nu}F_{\alpha\mu}F_{\beta\nu}
 +\int d^{4}x e^{\alpha\phi /M_{p}}(\Phi +k\sqrt{-g})
\frac{i}{2}\sum_{i}\overline{\Psi}_{i}
\left(\gamma^{a}e_{a}^{\mu}\overrightarrow{\nabla}^{(i)}_{\mu}-
\overleftarrow{\nabla}^{(i)}_{\mu}\gamma^{a}e_{a}^{\mu}\right)\Psi_{i}
\nonumber\\
&&-\int d^{4}xe^{\frac{3}{2}\alpha\phi /M_{p}} \left[(\Phi
+h_{E}\sqrt{-g})\mu_{E}\overline{E}E +(\Phi
+h_{N}\sqrt{-g})\mu_{N}\overline{N}N \right]. \label{totaction}
\end{eqnarray}
Here $\Psi_{i}$ ($i=N,E$) is the general notation for the
 fermion fields $N$ and $E$;
$F_{\alpha\beta}=\partial_{\alpha}A_{\beta}-\partial_{\beta}A_{\alpha}$;
$\mu_{N}$ and $\mu_{E}$ are the mass parameters;
$\overrightarrow{\nabla}^{(N)}_{\mu}=\vec{\partial}+
\frac{1}{2}\omega_{\mu}^{cd}\sigma_{cd}$,
$\overrightarrow{\nabla}^{(E)}_{\mu}=\vec{\partial}+
\frac{1}{2}\omega_{\mu}^{cd}\sigma_{cd}+ieA_{\mu}$;
 $R(\omega ,V)
=e^{a\mu}e^{b\nu}R_{\mu\nu ab}(\omega)$ is the scalar curvature;
$e_{a}^{\mu}$ and $\omega_{\mu}^{ab}$ are the vierbein and
spin-connection; $g^{\mu\nu}=e^{\mu}_{a}e^{\nu}_{b}\eta^{ab}$ and
$R_{\mu\nu ab}(\omega)=\partial _{mu}\omega_{\nu
ab}+\omega^{c}_{\mu a}\omega_{\nu cb}-(\mu \leftrightarrow\nu)$.
$V_{1}$ and $V_2$ are constants with the dimensionality
$(mass)^4$. When Higgs field is included into the model then
$V_{1}$ and $V_2$ turn into functions of the Higgs field. As we
will see later , in the Einstein frame $V_{1}$, $V_2$  and
$e^{\alpha\phi /M_{p}}$ enter in the effective potential of the
scalar sector. Constants $b, k, h_{N}, h_{E}$  are non specified
dimensionless real parameters;
 $\alpha$ is a real parameter
which we take to be positive.

The action (\ref{totaction}) is invariant under the global scale
transformations:
\begin{eqnarray}
    &&e_{\mu}^{a}\rightarrow e^{\theta /2}e_{\mu}^{a}, \quad
\omega^{\mu}_{ab}\rightarrow \omega^{\mu}_{ab}, \quad
\varphi_{a}\rightarrow \lambda_{ab}\varphi_{b}\quad
A_{\alpha}\rightarrow A_{\alpha}, \quad
 \phi\rightarrow
\phi-\frac{M_{p}}{\alpha}\theta \nonumber\\
&& \Psi_{i}\rightarrow e^{-\theta /4}\Psi_{i}, \quad
\overline{\Psi}_{i}\rightarrow e^{-\theta /4} \overline{\Psi}_{i},
\nonumber
\\
 &&\text{where} \quad \theta =const, \quad \lambda_{ab}=const \quad
 \text{and}
 \quad \det(\lambda_{ab})=e^{2\theta},
\label{stferm}
\end{eqnarray}
when the measure fields $\varphi_{a}$ are used for definition of
the measure $\Phi$, as in Sec.3.1. If the definition (\ref{Aabg})
of Sec.3.2 is used then the scale transformation of the totally
antisymmetric three index potential $A_{\alpha\beta\gamma}$ should
be: $A_{\alpha\beta\gamma}\rightarrow
e^{2\theta}A_{\alpha\beta\gamma}$.

Variation of the measure fields $\varphi_{a}$ (in the model with
the definition (\ref{Phi})  or $A_{\alpha\beta\gamma}$ (in the
model (\ref{Aabg})) yields Eq.(\ref{varphi}) where $L_{1}$ is now
defined, according to Eq.(\ref{S}), as the part of the integrand
of the action (\ref{totaction}) coupled to the measure $\Phi$. The
appearance of the integration constant $sM^{4}$ in
Eq.(\ref{varphi}) spontaneously breaks the global scale invariance
(\ref{stferm}).

Except for the $A_{\mu}$ equation, all other equations of motion
resulting from (\ref{totaction}) in the first order formalism
contain terms proportional to $\partial_{\mu}\zeta$ that makes the
space-time non-Riemannian and equations of motion - non canonical.
However, with the new set of variables ($\phi$ and $A_{\mu}$
remain unchanged)
\begin{equation}
\tilde{e}_{a\mu}=e^{\frac{1}{2}\alpha\phi/M_{p}}(\zeta
+b)^{1/2}e_{a\mu}, \quad
\tilde{g}_{\mu\nu}=e^{\alpha\phi/M_{p}}(\zeta +b)g_{\mu\nu}, \quad
\Psi^{\prime}_{i}=e^{-\frac{1}{4}\alpha\phi/M_{p}} \frac{(\zeta
+k)^{1/2}}{(\zeta +b)^{3/4}}\Psi_{i} , \quad i=N,E \label{ctferm}
\end{equation}
which we call the Einstein frame,
 the spin-connections become those of the
Einstein-Cartan space-time. Since $\tilde{e}_{a\mu}$,
$\tilde{g}_{\mu\nu}$, $N^{\prime}$ and $E^{\prime}$ are invariant
under the scale transformations (\ref{stferm}), spontaneous
breaking of the scale symmetry (by means of Eq.(\ref{varphi})) is
reduced in the new variables to the {\it spontaneous breaking of
the shift symmetry}\cite{Carroll}
\begin{equation}
\phi\rightarrow\phi +const. \label{phiconst}
\end{equation}
Notice that the Goldstone theorem generically is not applicable in
this theory (see the second reference in Ref.\cite{G1})). The
reason is the following. In fact, the scale symmetry
(\ref{stferm}) leads to a conserved dilatation current $j^{\mu}$.
However, for example in the spatially flat FRW universe the
spatial components of the current $j^{i}$ behave as $j^{i}\propto
M^4x^i$ as $|x^i|\rightarrow\infty$. Due to this anomalous
behavior at infinity, there is a flux of the current leaking to
infinity, which causes the non conservation of the dilatation
charge. The absence of the latter implies that one of the
conditions necessary for the Goldstone theorem is missing. The non
conservation of the dilatation charge is similar to the well known
effect of instantons in QCD where singular behavior in the spatial
infinity leads to the absence of the Goldstone boson associated to
the $U(1)$ symmetry.

\section{TMT structure, cosmological constant problem and  avoidance of the Weinberg
no-go theorem}

One can show\cite{GK4},\cite{HIS} that in the absence of massive
fermions the constraint determines $\zeta$ as the function of
$\phi$ alone:
\begin{equation}
\zeta =\zeta_{0}(\phi)\equiv b-\frac{2V_{2}}
{V_{1}+M^{4}e^{-2\alpha\phi/M_{p}}}, \label{zeta-without-ferm}
 \end{equation}
Note that the electromagnetic field does not enter in the
constraint and therefore the presence of the electromagnetic field
does not affect the value of $\zeta_{0}$.

 The effective potential of
the scalar field $\phi$ produced in the Einstein frame reads
\begin{equation}
V_{eff}^{(0)}(\phi)\equiv
V_{eff}(\phi;\zeta_{0})|_{\overline{\psi^{\prime}}\psi^{\prime}=0}
=\frac{[V_{1}+M^{4}e^{-2\alpha\phi/M_{p}}]^{2}}
{4[b\left(V_{1}+M^{4}e^{-2\alpha\phi/M_{p}}\right)-V_{2}]}
\label{Veffvac}
\end{equation}
and the $\phi$-equation  has the standard form
\begin{equation}
(-\tilde{g})^{-1/2}\partial_{\mu}
(\sqrt{-\tilde{g}}\tilde{g}^{\mu\nu}\partial_{\nu}\phi)
+V^{(0)\prime}_{eff}(\phi)=0, \label{eq-phief-without-ferm}
\end{equation}
where prime sets derivative with respect to $\phi$.

 The mechanism of the appearance of the effective potential (\ref{Veffvac})
 is very interesting and exhibits the main features of our TMT
 model\footnote{The particular case of this model
with $b=0$ and $V_{2}<0$ was studied in Ref.\cite{G1}. The
application of the TMT model with explicitly broken global scale
symmetry to the quintessential inflation scenario was discussed in
Ref.\cite{K}. A particular case of the model (\ref{totaction})
without explicit potentials, i.e. $V_{1}=V_{2}=0$, has been
studied in Ref.\cite{GK4}.}.
 In fact, the reasons of the transformation of the
 prepotentials $V_{1}e^{2\alpha\phi/M_{p}}$
 and $V_{2}e^{2\alpha\phi/M_{p}}$, coming in the original action
 (\ref{totaction}),  into the effective potential (\ref{Veffvac})
 are the following:

a) Transformation to the Einstein frame;

b)  Spontaneous breakdown of the global scale symmetry which in
the Einstein frame is reduced to the spontaneously broken shift
symmetry (\ref{phiconst});

c) The constraint which in the absence of fermions case takes the
form (\ref{zeta-without-ferm}).

It turns out that the gravitational equations  in the absence of
fermions case become the standard Einstein equations of the model
where the electromagnetic field and the minimally coupled scalar
field $\phi$ with the potential $V_{eff}^{(0)}(\phi)$ are the
sources of gravity.

The most remarkable feature of the effective potential
(\ref{Veffvac}) is that it is proportional to the square of
$V_{1}+ M^{4}e^{-2\alpha\phi/M_{p}}$. Due to this, as $V_{1}<0$
and $bV_1-V_{2}>0$, {\it the effective potential has a minimum
where it equals zero automatically}, without any further tuning of
the parameters $V_{1}$ and $V_{2}$. This occurs in the process of
evolution of the field $\phi$ at the value of $\phi =\phi_{0}$
where
\begin{equation}
V_{1}+ M^{4}e^{-2\alpha\phi_{0}/M_{p}}=0 \label{Veff=0}.
\end{equation}
This means that the universe evolves into the state with zero
cosmological constant without tuning  parameters of the model.

 If such type of the structure for the scalar field potential in a
usual (non TMT) model would be chosen "by hand" it would be a sort
of fine tuning. But in our TMT model it is not the starting point,
{\it it is rather a result} obtained in the Einstein frame of TMT
models with spontaneously broken global scale symmetry including
the shift symmetry $\phi\rightarrow \phi +const$.  The same effect
can be obtained in more general models including for example the
Higgs field as well\cite{HIS}. Note that the assumption of scale
invariance is not necessary for the effect of appearance of the
perfect square in the effective potential in the Einstein frame
and therefore for the described mechanism of disappearance of the
cosmological constant, see Refs.\cite{GK2}-\cite{G1}.

On the first glance this effect contradicts the no-go Weinberg
theorem\cite{Weinberg1} which states that there cannot exist a
field theory model where the cosmological constant is zero without
fine tuning. Recall that one of the basic assumptions of this
no-go theorem is that all fields in the vacuum must be constant.
However, this is not the case in TMT.   In fact, in the vacuum
determined by Eq.(\ref{Veff=0}) the scalar field
$\zeta\equiv\frac{\Phi}{\sqrt{-g}}$ is non zero, see
Eq.(\ref{zeta-without-ferm}). The latter is possible only if all
the $\varphi_{a}$ ($a=1,2,3,4$) fields (in the definition of
$\Phi$ by means of Eq.(\ref{Phi})) or the 3-index potential
$A_{\alpha\beta\gamma}$ (when using the definition of $\Phi$ by
means of Eq.(\ref{Aabg})) have non vanishing space-time gradients.
Moreover, exactly in the vacuum $\phi =\phi_{0}$ the scalar field
$\zeta$ has a singularity. However, in the conformal Einstein
frame all physical quantities are well defined and this
singularity manifests itself only in the vanishing of the vacuum
energy density. We conclude therefore that the Weinberg
theorem\cite{Weinberg1} is not applicable in the context of the
TMT models studied here. In fact, the possibility of such type of
situation was suspected by S. Weinberg in the footnote 8 of his
review\cite{Weinberg1} where he pointed out that when using a
3-index potential with non constant vacuum expectation value, his
theorem does not apply.

When considering more realistic models, the pre-potentials $V_{1}$
and $V_{2}$ turn into functions of the Higgs field. If $V_{1}$ and
$V_{2}$ include for example both Higgs mass terms and quartic
Higgs self-interactions, then there could be vacua with zero
cosmological constant disconnected from each other (again without
fine tuning). This is an explicit realization of the "Multiple
Point Principle" proposal\cite{Nielsen} which is based on the idea
that if there is a mechanism that sets a certain state to have a
zero cosmological constant  then the same mechanism  may act also
in  other  field configurations with the same result.

\section{Phenomenological output of TMT}

Here we list some of the most interesting results of TMT:\\
1. Restoration of GR as local fermion energy density is much
larger than the scalar dark energy in the space-time region
occupied by the fermion\cite{GK4}-\cite{GK6}:

 a) restoration of the Einstein equations;

 b) decoupling of the scalar field $\phi$ from fermionic matter;

 c) some ideas concerning the nature of three fermion generations
 are also discussed.\\
2. As local fermion energy density has the order of magnitude
close to that of the scalar dark energy, the fermion may be in an
exotic state\cite{GK6}: gas of nonrelativistic neutrinos in such a
state behaves as dark energy.\\
3. Power law inflation of the early universe, driving by the
scalar sector (inflaton $\phi$ and the Higgs field $\upsilon$),
consists of two stages\cite{HIS}. In the second stage, starting
about 50 e-folding before the end of inflation, $\upsilon$
performs quickly
damping oscillations around $\upsilon =0$.\\
4. Soon after the end of inflation, a transition to the gauge
symmetry broken phase, which at the same time is a true vacuum
state with zero energy density, is realized\cite{HIS}. This
transition is accompanied with the Higgs field oscillations that
should serve as
an effective mechanism of reheating. \\
5. Zero vacuum energy and spontaneous breakdown of the gauge
symmetry occur without any fine tuning and without tachyonic mass
term in the action\cite{HIS}.\\
6. Other scenario is that of early inflation driven by $R^2$ term,
followed by transition to slowly accelerated phase\cite{GKatz}.\\
 7. For the late time universe a possibility of
superaccelerated expansion of the universe is shown without
phantom field in the
action\cite{HIS}. \\
8. Smallness of the present cosmological constant may be obtained
by a see-saw mechanism\cite{GK6},\cite{HIS}.

\section{Problems of TMT Quantization. Preliminary discussion}

So far we have discussed effects only in classical two measures
field theory. However quantization of TMT as well as influence of
quantum effects on the processes explored might have a crucial
role. We summarize here some ideas and speculations which gives us
a hope that quantum effects can keep and even strengthen the main
results displayed in classical TMT.

1) Recall first two fundamental facts of TMT as a classical field
theory: (a) The measure degrees of freedom appear in the equations
of motion only via the scalar $\zeta$; (b) The scalar $\zeta$ is
determined by  the constraint which is nothing but a consistency
condition of the equations of motion. Therefore the constraint
plays a key role in TMT. Note however that if we were ignore the
gravity from the very beginning in the action (\ref{totaction})
then instead of the constraint we would obtain a very different
equation, i.e. in such a case we would deal with a different
theory. This notion shows that the gravity and matter intertwined
in TMT in a much more complicated manner than in GR. Hence
introducing the new measure of integration $\Phi$ we have to
expect that the quantization of TMT may be a complicated enough
problem. Nevertheless we would like here to point out that in the
light of the recently proposed idea of Ref.\cite{Giddings}, the
incorporation of four scalar fields $\varphi_a$  together with the
scalar density $\Phi$, Eq.(\ref{Phi}), (which in our case are the
measure fields and the new measure of integration respectively),
is  a possible way to define local observables in the local
quantum field theory approach to quantum gravity. We regard this
result as an indication that the effective gravity $+$ matter
field theory has to contain the new measure of integration $\Phi$
as it is in TMT. In fact in our motivation of TMT from the point
of view of the brane world scenario, Eq.(\ref{NG}), the measure
fields $\varphi_a$ have a meaning of the coordinates of the
embedding 3-brane. This supports both the idea of the measure
fields as physical coordinates\cite{Giddings} and of the structure
of TMT expressed by Eq.(\ref{S}).

2)  The assumption that the measure fields $\varphi_a$ (or
$A_{\alpha\beta\gamma}$) appear in the action (\ref{S}) {\it only}
via the measure of integration $\Phi$, has a key role in the TMT
results and in particular for the resolution of the old
cosmological constant problem. In principle one can think of
breakdown of such a structure by quantum corrections. However,
fortunately there exists an infinite dimensional symmetry
(\ref{inf-dim-symm})\, (and its analog in the case (\ref{Aabg}))
which, as we hope, is able to protect the postulated structure of
the action from a deformation caused by quantum corrections or at
least to suppress such a quantum anomaly in significant degree.
Therefore one can hope that the proposed resolution of the old
cosmological constant problem holds in the quantized TMT as well.

\section{Acknowledgments}
  We thank L. Amendola, S. Ansoldi, R. Barbieri, J. Bekenstein, A.
Buchel, A. Dolgov,  S.B. Giddings, P. Gondolo, J.B. Hartle, F.W.
Hehl, B-L. Hu, P.Q. Hung, D. Kazanas, O. Katz, D. Marolf, D.G.
McKeon, J. Morris, V. Miransky, Y. Ne'eman, H. Nielsen, Y.Jack Ng,
H. Nishino, E. Nissimov,  S. Pacheva, L. Parker, R. Peccei, M.
Pietroni, S. Rajpoot, R. Rosenfeld,  V. Rubakov, E. Spallucci, A.
Starobinsky, G. 't Hooft, A. Vilenkin, S. Wetterich and  A. Zee
for helpful conversations on different stages of this research.

\end{document}